\documentclass{raa}            

\usepackage{graphicx,times}             

\begin{document}

  \title{A note on the puzzling spindown behavior of the Galactic center magnetar SGR J1745$-$2900
}

   \volnopage{Vol.0 (200x) No.0, 000--000}      
   \setcounter{page}{1}          

     \author{H. Tong
        }

   \institute{Xinjiang Astronomical Observatory, Chinese Academy of Sciences, Urumqi, Xinjiang 830011,
    China; {\it tonghao@xao.ac.cn}\\
                  }

   \date{Received~~2012 month day; accepted~~2012~~month day}

\abstract{SGR J1745$-$2900 is a magnetar near the Galactic center.
X-ray observations of this source found a decreasing X-ray
luminosity accompanied by an enhanced spindown rate. 
This negative correlation between X-ray luminosity and spindown rate is hard to understand. 
The wind braking model of magnetars is employed to explain this puzzling spindown behavior.
During the release of magnetic energy of magnetars, a system of particles may be generated.
Some of these particles remain trapped in the magnetosphere and may contribute to the X-ray luminosity.
The rest of the particles can flow out and take away the rotational energy of 
the central neutron star.
A smaller polar cap angle will cause the decrease of X-ray luminosity and enhanced spindown rate of
SGR J1745$-$2900. This magnetar is expected to have a maximum spindown rate shortly.
 }

   \authorrunning{H. Tong}            
   \titlerunning{On the puzzling spindown behavior of magnetar SGR J1745$-$2900}  

   \maketitle

%
%

\section{Introduction}

Magnetars are a special kind of pulsars. They are assumed to be neutron stars whose radiative activities are
powered by their magnetic energy (Duncan \& Thompson 1992). 
Both the radiative and timing properties of magnetars vary with time. During the outburst of a magnetar,
the star's X-ray luminosity increase significantly and then decay gradually (Rea \& Esposito 2011).
The outburst may also accompanied by timing events, e.g., spin-up glitch (Kaspi et al. 2003),
or spin-down glitch (Archibald et al. 2013; Tong 2014), and/or period derivative changes.
Many magnetars show different degrees of period derivative variations (Woods et al. 2007; Dib \& Kaspi 2014). 
The timing variabilities imply that the
magnetic dipole braking in vacuum is a poor approximation in the case of magnetars
and magnetars may be wind baking (Tong et al. 2013).

The twisted magnetosphere model tried to understand the radiative and timing behaviors of magnetars
using an untwisting neutron star magnetosphere (Thompson et al. 2002; Beloborodov 2009).
During the outburst a magnetar is expected to have a decreasing  X-ray luminosity and
decreasing spindown rate. 
However, the timing and flux evolution of the Galactic center magnetar SGR J1745$-$2900
showed a negative correlation between X-ray luminosity and spindown rate (Kaspi et al. 2014).
During the nearly four months observations, the magnetars's X-ray luminosity decreased by a
factor of two. While, the spindown rate has increased by a factor of 2.6.
And the spindown rate is still increasing. 
Kaspi et al. (2014) discussed changes in the open field line regions.
However, there is no quantitative calculation at present.
The structure of open and closed field line regions of magnetar magnetosphere has been calculated 
in the wind braking model (Tong et al. 2013). The puzzling spindown behavior
of the Galactic center magnetar may be understandable in the wind braking model.
The spindown rate of SGR J1745$-$2900 may contain some contribution from the 
Galatic center black hole (Rea et al. 2013). Therefore, understanding its 
spindown behavior is very important. 

In the wind braking model of magnetars (Tong et al. 2013),
a particle outflow is generated during the decay of the magnetic field.
Some of these particles may remain confined in the magnetosphere and contribute to the
X-ray luminosity. The rest of the particles can flow out to infinity (i.e., wind),
thus dominate the spindown of the magnetar. 
The particle outflow may be mainly confined in a specific polar cap of the central neutron star.
For a smaller polar cap angle, it will result in a smaller X-ray luminosity and larger spindown rate.
This may explain the puzzling spindown behavior of SGR J1745$-$2900.

Calculations in the wind braking model for SGR J1745$-$2900 are given in Section 2.
Discussions and conclusions are presented in Section 3.

\section{Wind braking model for the spindown behavior of SGR J1745$-$2900}

\subsection{Qualitative descriptions}

\begin{figure}[!t]
 \centering
\includegraphics{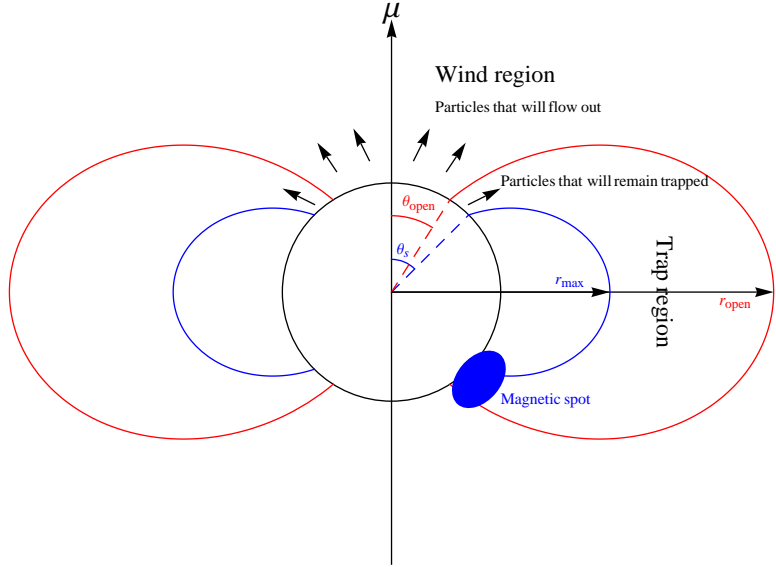}
\caption{The magnetosphere in the presence of a particle wind, for illustrative use only.}
\label{fig-magnetosphere}
\end{figure}

Figure \ref{fig-magnetosphere}
shows the magnetar magnetosphere in the case of wind braking. The decay of the star's
magnetic energy may result in a bunch of particles, with total particle luminosity $L_{\rm p}$.
The particle luminosity
may be similar to the magnetar's X-ray luminosity\footnote{The particle energy can be converted to
either soft X-ray or hard X-ray luminosity (Thompson et al. 2002; Tong et al. 2010; Beloborodov 2013). 
For SGR J1745$-$2900, its soft X-ray luminosity
dominates the electromagnetic energy output (Kaspi et al. 2014).}
($L_{\rm p} \sim L_{\rm x} \sim 10^{35} \,\rm erg\, s^{-1}$,
Duncan 2000). There exist a maximum length $r_{\rm max}$ for the coupling between the neutron star
and its magnetosphere.
The corresponding polar cap angle is $\theta_{s}$.
From quasi-periodic oscillation observations in magnetars, $\theta_{\rm s}$ is about $0.05$
(Timokhin et al. 2008; Watts 2011; Tong et al. 2013).
All particles in this polar cap will try to flow out to infinity.
Considering the presence of large scale dipole magnetic field, some of these particle will be trapped.
The opening radius $r_{\rm open}$ is defined as
the radius where the particle kinetic energy density equals the local magnetic energy density
(Harding et al. 1999; Thompson et al. 2000; Tong et al. 2013).
Assuming uniform distribution of particles across the polar cap, $r_{\rm open}$ is (Equation (20) in Tong et al. 2013)
\begin{equation}\label{ropen}
r_{\rm open} = 7\times 10^9 \, b_0^{2/3} L_{\rm p,35}^{-1/3} (\theta_{\rm s}/0.05)^{2/3} \,\rm cm,
\end{equation}
where $b_0$ is the surface dipole field (at the magnetic pole)
in units of the quantum critical field ($4.4\times 10^{13}\rm\, G$), $L_{\rm p,35}$ is the
total particle luminosity in units of $10^{35} \,\rm erg \,s^{-1}$. The polar cap angle of the
magnetic field line whose maximum radius equals $r_{\rm open}$ is $\theta_{\rm open} =(r_0/r_{\rm open})^{1/2}$
(where $r_0$ is the neutron star radius, it is taken as $10^6\,\rm cm$ during numerical calculations).

For particles in the polar cap region with $\theta<\theta_{\rm open}$,
 they can flow out to infinity and carry away the star's rotational energy (dubbed as the ``wind region'').
 This particle component may dominate the spindown behavior
of the central neutron star. It is denoted as $L_{\rm w}$ (the wind luminosity). 
While, for particles in the polar cap region with
$\theta_{\rm open} < \theta < \theta_{\rm s}$, they will be trapped by the dipole magnetic field
(dubbed as the ``trap region'').
These particles may finally contribute to the X-ray luminosity of the magnetar.
If near one foot of the field lines, there is a domain with higher magnetic field strength than
the surrounding area (``magnetic spot'', in analogy with the sun spot), a hot spot may be formed.
The pulse profile of SGR J1745$-$2900 has multipeaks (Kaspi et al. 2014). This may due to the presence of
multipole field (i.e., a magnetic spot).

The total particle luminosity is equal  to the sum of wind luminosity and X-ray luminosity\footnote{There may
be other sources of X-ray luminosity, e.g., internal heating (Vigano et al. 2013).
As long as the contribution from outflowing particles is comparable to those sources (e.g., they contribute equally),
the description in this section is still valid.}: $L_{\rm p} = L_{\rm w} + L_{\rm x}$.
Observationally, some magnetars can have a nearly constant X-ray luminosity early during the outburst (lasts for about 100 days,
Rea \& Esposito 2011). Then it is also possible that the total particle luminosity
can maintain nearly constant for about 100 days. For a constant particle luminosity $L_{\rm p}$, if the polar cap radius
$\theta_{\rm s}$ changes, it will cause a negative correlation between X-ray luminosity and wind luminosity (i.e., spindown rate).
The change in polar cap radius may be accompanied/triggered by short burst. For SGR J1745$-$2900, it has a short burst on MJD 56450
(Kennea et al. 2013; Kaspi et al. 2014).
After the short burst, some local multipole field may annihilate (which may correspond to the disappearance of one pulse
profile peak, Kaspi et al. 2014).
The coupling length scale between the neutron star and the magnetosphere $r_{\rm max}$ will increase (due to a simplified 
magnetic field geometry).
The polar cap angle $\theta_{\rm s}$ will decrease. 
This may correspond to the decreasing X-ray luminosity and enhanced spindown rate of SGR J1745$-$2900
(Kaspi et al. 2014). Furthermore, the space density of the particles may increase (due to
a smaller polar cap area). This will result in a harder spectrum (Kaspi et al. 2014).
All these aspects are qualitatively in agreement with the observations of SGR J1745$-$2900, especially
the decreasing X-ray luminosity and the enhanced spindown rate.

\subsection{Calculations for a constant particle luminosity}

According to the wind braking model of magnetars (Tong et al. 2013), 
the star's rotational energy loss rate due to the presence of a particle wind is 
(Equation (23) in Tong et al. 2013)
\begin{equation}\label{Edotw}
 \dot{E}_{\rm w} = \dot{E}_{\rm d} \left( \frac{L_{\rm w}}{\dot{E}_{\rm d}} \right)^{1/2},
\end{equation}
where $\dot{E}_{\rm w}$ is the rotational energy carried away per unit time by the particle wind, 
and $\dot{E}_{\rm d}$ is the corresponding rotational energy loss rate for the simple dipole spindown case. 
The presence of particle wind results in an enhancement of the rotational energy loss rate. 
The wind luminosity $L_{\rm w}$ depends on the total particle luminosity and the polar cap angle
(Equation (22) in Tong et al. 2013)
\begin{equation}
 L_{\rm w} = 6\times 10^{33} b_0^{-2/3} L_{\rm p, 35}^{4/3} (\theta_{\rm s}/0.05)^{-8/3} \,\rm erg \ s^{-1}. 
\end{equation}
Therefore, in the wind braking model of magnetars, the star's polar surface dipole field is
(Equation (24) in Tong et al. 2013):
\begin{equation}\label{B0}
B_0 = 3.3\times 10^{32} \left( \frac{\dot{P}}{P} \right)^{3/2} L_{\rm p, 35}^{-1}
(\theta_{\rm s}/0.05)^2 \,\rm G,
\end{equation}
where $B_0$ is the polar surface dipole field, $P$ and $\dot{P}$
are the star's rotation period and period derivative, respectively. For a short duration (e.g., 100 days),
the star's surface dipole field and rotation period change very little. Assuming a constant
total particle luminosity, then the period derivative is proportional to
\begin{equation}
\dot{P} \propto (\theta_{\rm s}/0.05)^{-4/3}.
\end{equation}
Observationally, the spindown rate of SGR J1745$-$2900 increased by a factor of $2.6$ (Kaspi et al. 2014).
This requires that the polar cap angle is about $0.5$ times the initial value:
$\theta_{\rm s,f} =0.5 \theta_{\rm s,i}$ (where $\theta_{\rm s,i}$ and $\theta_{\rm s,f}$ are the
initial and final polar cap angle during the observations, respectively).
The wind luminosity is related with the total particle luminosity as
\begin{equation}
L_{\rm w} = L_{\rm p} \frac{\theta_{\rm open}^2}{\theta_{\rm s}^2},
\end{equation}
since they are proportional to the corresponding polar cap area.
The X-ray luminosity is
\begin{equation}\label{Lx and Lp}
L_{\rm x} = L_{\rm p} -L_{\rm w} =L_{\rm p} \left( 1- \frac{\theta_{\rm open}^2}{\theta_{\rm s}^2} \right).
\end{equation}
The X-ray luminosity of SGR J1745$-$2900 decreased by a factor of $2$ during the observations (Kaspi et al. 2014).
This requires that
\begin{equation}\label{Lx ratio}
\frac{L_{\rm x,f}}{L_{\rm x,i}} =
\frac{1-(\theta_{\rm open}^2/\theta_{\rm s}^2)_{\rm f}}{1-(\theta_{\rm open}^2/\theta_{\rm s}^2)_{\rm i}} =0.5,
\end{equation}
where $L_{\rm x,i}$ and $L_{\rm x,f}$ are the initial and final X-ray luminosities, respectively.
From Equation (\ref{ropen}), $\theta_{\rm open}^2 \propto 1/r_{\rm open} \propto \theta_{\rm s}^{-2/3}$.
Then $\theta_{\rm open}^2/\theta_{\rm s}^2 \propto \theta_{\rm s}^{-8/3}$. Using the above timing results,
Equation (\ref{Lx ratio}) can be rewritten as
\begin{equation}
\frac{1-2.6^2 \,(\theta_{\rm open}^2/\theta_{\rm s}^2)_{\rm i}}{1 -(\theta_{\rm open}^2/\theta_{\rm s}^2)_{\rm i}} =0.5.
\end{equation}
Therefore, the two polar cap angles are related with each other as
\begin{equation}\label{thetaopen and thetas}
(\theta_{\rm open}^2/\theta_{\rm s}^2)_{\rm i} =0.08,
\end{equation}
where a subscript $\rm i$ means the initial value. From Equation (\ref{Lx and Lp}) and using flux observations of SGR J1745$-$2900
(Kaspi et al. 2014), the total particle luminosity in the wind braking model is:
$L_{\rm p} =1.7\times 10^{35} \,\rm erg \, s^{-1}$ (assuming a flux of $2\times 10^{-11} \,\rm erg \, s^{-1} \, cm^{-2}$
and a distance of $8\,\rm kpc$). Solving Equation (\ref{B0}) and (\ref{thetaopen and thetas}) together,
the surface dipole field and initial polar cap angle can be obtained:
$B_0 = 1.8\times 10^{14} \,\rm G$ and $\theta_{\rm s,i} =0.033$. $B_0$ is lower than the
characteristic magnetic field when assuming magnetic dipole braking in vacuum
(at the magnetic pole, which is $3.1\times 10^{14} \,\rm G$ using the first ehpemeris in Kaspi et al. 2014). 
The corresponding opening radius is (equation (\ref{ropen})) $1.1\times 10^{10} \,\rm cm$. 
It is slightly smaller than the light cylinder radius $1.8\times 10^{10} \,\rm cm$. 
This is because the particle wind will ``comb out'' magnetic field lines, contribute
to the rotational energy loss rate and a lower surface dipole field is required.
For SGR J1745$-$2900, its particle wind is a little bit stronger than its dipolar radiation. 
$\theta_{\rm s,i}$ is similar to the estimated value using quasi-periodic oscillation observations (which is about 0.05).
This is the first time that the polar cap angle ($\theta_{\rm s}$) of magnetar wind can be determined.

Therefore, in the wind braking model, SGR J1745$-$2900 is a magnetar with surface dipole field
$B_0 =1.8\times 10^{14} \,\rm G$ and total particle luminosity $L_{\rm p} =1.7\times 10^{35} \,\rm erg \,s^{-1}$.
Its initial polar cap angle is $\theta_{\rm s,i} =0.033$. During the observation, its polar cap angle
changes to $0.5$ times the initial value. This causes the decrease of X-ray luminosity and enhanced
spindown rate (Kaspi et al. 2014). 

\subsection{Calculations when the particle luminosity decreases as the the X-ray luminosity}

In the above calculations, a constant particle luminosity is assumed and the X-ray luminosity
is dominated by particles in the closed field line regions.
Another extreme case is that both particles in the open field line regions and the closed field 
line regions may contribute to the X-ray luminosity. Then a model independent estiamtion 
for the particle luminosity is $L_{\rm p} = L_{\rm x}$ (Duncan 2000; Tong et al. 2013). 
During the outburst, a decreasing magnetic energy release rate will cause a decreasing 
X-ray luminosity (e.g., Thompson \& Duncan 1996; Beloborodov 2009). From Equation (\ref{B0}), 
in the present case, the star's spindown rate is proportional to 
\begin{equation}\label{PdotCaseII}
 \dot{P} \propto L_{\rm p}^{2/3} \theta_{\rm s}^{-4/3} = L_{\rm x}^{2/3} \theta_{\rm s}^{-4/3}. 
\end{equation}
For SGR J1745$-$2900, its X-ray luminosity decreased by a factor of two, while its spindown 
rate incresed by a factor of 2.6 during the observations (Kaspi et al. 2014). 
This requires that the final polar cap angle is smaller by 
factor of 0.35: $\theta_{\rm s,f}=0.35 \theta_{\rm s,i}$. The determination of the exact value
of the polar cap angle will require addiational assumptions. However, it will be different 
only in quantity from the above calculations. The physical reason for the negative 
correlation between the X-ray luminosity and the spindown rate is the same: a smaller polar cap 
opening angle will result in a higher spindown rate.

\section{Discussions and conclusions}

In pulsar studies, the magnetic dipole braking assumption is often employed. However, 
it assumes an orthogonal rotator in vacuum (Shapiro \& Teukolsky 1983). 
A real pulsar must have a magnetosphere. 
The magnetosphere of magnetars may be twisted compared with normal pulsars
(Thompson et al. 2002; Beloborodov 2009). 
However, the twisted magnetosphere model does not consider the rotation of the central 
neutron star (Thompson et al. 2002). All the field lines are closed in the twisted magnetosphere model. 
There is no open field line (i.e., no polar cap). Considering current modeling of normal pulsar spindown 
(wind braking, Xu \& Qiao 2001; Li et al. 2014), magnetars may also be spun down by a particle wind
(Tong et al. 2013). 
The particle wind luminosity (powered by the magnetic energy) may be much higher than the rotational energy loss rate.
This will make wind braking of magnetars qualitatively different from wind braking of normal pulsars 
(Tong et al. 2013). 
The particle luminosity can also vary dramatically as that of the X-ray luminosity
(since they are both powered by the magnetic energy). 
This may explain why magnetars can have so many timing events (for a summary see Tong \& Xu 2014). 
Compared with magnetic dipole braking, the wind braking model considers the existence of a neutron star magnetosphere. 
Compared with the twisted magnetosphere model, the wind braking model considers the rotation of the central neutron star from the starting point.
The wind braking model of magnetars (Tong et al. 2013; and the calculations here) considered an aligned rotator, and
a uniform charge density over the polar cap.  
The spindown behavior is mainly determined by the total particle outflow. It is not very sensitive 
to the inclination angle and charge distribution.

Assuming the observed X-ray luminosity is solely due to other sources, it is unavoidable that a particle wind is
generated during the decay of the magnetic field (Thompson \& Duncan 1996; Beloborodov \& Thompson 2007).
This particle wind will have a comparable luminosity to that of the X-ray emissions
$L_{\rm p} \sim L_{\rm x} \sim 10^{35} \,\rm erg \,s^{-1}$ (Thompson \& Duncan 1996; Duncan 2000; Tong et al. 2013).
Using Equation (\ref{B0}), the corresponding surface dipole field is
$B_0 \approx 6.8\times 10^{14} \,(\theta_{\rm s}/0.05)^2 \,\rm G$. It is similar to the surface dipole field
in the case of magnetic dipole braking. Only a small fraction
of the particles flows out to infinity. Most of them remain trapped in the magnetosphere.
They will collide with the stellar surface, scattering off X-ray photons etc
(Thompson et al. 2002; Beloborodov \& Thompson 2007; Tong et al. 2010). Another X-ray component is generated
which is comparable to the original X-ray component. Therefore, the outflowing particles must
contribute a significant fraction to the X-ray luminosity. Previous studies favor a magnetospheric origin for
the X-ray luminosity during the outburst (Beloborodov 2011).
The upper limit of quiescent X-ray luminosity may put strong constraint on the contribution from
other persistent energy sources (Mori et al. 2013). In summary, the assumption is reasonable that the X-ray luminosity
(at least half of it) may be dominated by the contribution of outflowing particles.

According to the numerical calculations, the total particle luminosity of SGR J1745$-$2900
is $L_{\rm p} =1.7\times 10^{35} \,\rm erg \,s^{-1}$. If all these particles can flow out, this corresponds
to the maximum spindown case $L_{\rm w} = L_{\rm p}$ (Tong et al. 2013, Section 3.2).
Therefore, SGR J1745$-$2900 has a maximum spindown rate.
Using Equation (31) in Tong et al. (2013), the maximum period derivative is
$\dot{P}_{\rm max} = 2.2 \times 10^{-11}$. It is about two times higher than the period derivative
in the second ephemeris of SGR J1745$-$2900 (Kaspi et al. 2014). Using the period second derivative
measurement (Kaspi et al. 2014), the time required to reach this maximum spindown state is
$(\dot{P}_{\rm max} -\dot{P})/\ddot{P} \approx 240 \,\rm days$. Therefore, 
SGR J1745$-$2900 may reach a state with maximum spindown rate in about one year. If the particle luminosity
decreases a little at that time, the correspond period derivative will also be a little smaller.
The X-ray luminosity during the maximum spindown state will be very small since a very small fraction
of particles is trapped. Other sources of X-ray luminosity
may only contribute a relatively small part of the X-ray luminosity (Mori et al. 2013).
From previous experiences of magnetar outburst (Rea \& Esposito 2011),
both the X-ray luminosity and spindown rate will decrease long after the outburst.

The timing behavior of magnetar Swift J1822.3$-$1606 is governed by the change of wind luminosity (Tong \& Xu 2013).
On the other hand, the timing behavior of magnetar SGR J1745$-$2900 may be dominated by the change of polar cap angle.
In general, both the particle luminosity and the polar cap angle vary with time after the outburst. This may
explain the different radiative and timing correlations in magnetars (Dib \& Kaspi 2014 and references therein).
Combined with timing studies of pulsars (Li et al. 2014 and references therein),
not only magnetars but also normal pulsars are wind braking (see Equation (\ref{Edotw})). 

One consequence of the wind braking model of magnetars is a magnetism-powered pulsar wind nebula
(Tong et al. 2013). There is one weak evidence (Younes et al. 2012). From the pulsar wind
nebulae observations in normal pulsars, the nebula luminosity
is only about $10^{-4}$ times the total particle luminosity\footnote{The particle luminosity
is equal to the rotational energy loss rate in the case of normal pulsars. } (Kargaltsev et al. 2013).
For SGR J1745$-$2900, its particle luminosity is about $10^{35} \,\rm erg \,s^{-1}$.
At a distance of $8\,\rm kpc$, with an X-ray efficiency of about $10^{-4}$, it is unlikely that
the nebula can be observed using current telescopes (Kargaltsev et al. 2013). Furthermore, the particle wind
in the case of magnetars may only exist for several years (the same duration as the outburst).
And its luminosity also decreases with time.
This will make its detection more difficult. Current non-detections
of wind nebula are not constraining (e.g., Archibald et al. 2013; Scholz et al. 2014).

In conclusion, in the wind braking model of magnetars,
change of polar cap angle may cause a negative correlation between the X-ray luminosity and
the spindown rate.
A polar cap angle $0.5$ times the initial value will explain the decrease of X-ray luminosity (by a factor of two)
and enhancement of spindown rate (by a factor of 2.6) of SGR J1745$-$2900. SGR J1745$-$2900 is expected to reach
a state with maximum spindown rate shortly.

\appendix

\section{Comparison with later observation}

After this paper was submitted and put on the arXiv, Lynch et al. (2014) reported their updated 
timing of SGR J1745$-$2900. According to their polynomial fitting, the frequency derivatives of this source 
are $\dot{f}=-1.24 \times 10^{-12} \,\rm Hz \,s^{-1}$, $\ddot{f}=-1.05 \times 10^{-19} \,\rm Hz \,s^{-2}$, 
and $\stackrel{...}{f}=6.7 \times10^{-27} \,\rm Hz \,s^{-3}$ (Table 1 in Lynch et al. 2014). 
Therefore, the spindown rate will first increase (due to a negative $\ddot{f}$), reaching a maximum then decrease
(due to a positive $\stackrel{...}{f}$). This is qualitatively consistent with our model calculations. 
However, the polynomial fitting (twelve frequency derivatives are employed by Lynch et al. 2014) absorbs 
all the physical changes and has no predictive power. The spindown rate as a function of time is required 
in order to determine the origin of torque variations.

\section*{Acknowledgments}
The author would like to thank C.Wang, J.L.Han and R.X.Xu for discussions.
H.Tong is supported by Xinjiang Bairen project, NSFC (11103021), West Light Foundation of CAS (LHXZ201201),
and Qing Cu Hui of CAS, and  973 Program (2015CB857100).

\label{lastpage}


\begin{thebibliography}{99}

\bibitem{Archibald2013}
Archibald, R. F., Kaspi, V. M., Ng, C. Y., et al. 2013, Nature, 497, 591

\bibitem{Beloborodov (2007)}
Beloborodov, A. M., \& Thompson, C. 2007, ApJ, 657, 967

\bibitem{Beloborodov2009}
Beloborodov, A. M. 2009, ApJ, 703, 1044

\bibitem{Beloborodov2011}
Beloborodov, A. M. 2011, in N. Rea and D.F. Torres (eds.), High-Energy Emission from Pulsars and their Systems,
299 (arXiv:1008.4388)

\bibitem{Beloborodov2013}
Beloborodov, A. 2013, 762, 13

\bibitem{Dib2014}
Dib, R., \& Kaspi, V. M. 2014, ApJ, 784, 37

\bibitem{DT1992}
Duncan, R. C., \& Thompson, C. 1992, ApJ, 392, L9

\bibitem{Duncan2000}
Duncan, R. C. 2000, AIP conference proceedings, 526, 830

\bibitem{Harding 1999}
Harding , A. K., Contopoulos, I., \& Kazanas, D. 1999, ApJ, 525, L125

\bibitem{Kargaltsev2010}
Kargaltsev, O., B. Rangelov, \& Pavlov, G. G. 2013, arXiv:1305.2552

\bibitem{Kaspi2003}
Kaspi, V. M., Gavriil, F. P., \& Woods, P. M. 2003, ApJ, 588, L93

\bibitem{Kaspi2014}
Kaspi, V. M., Archibald, R. F., Bhalerao, V., et al. 2014, ApJ, 786, 84	

\bibitem{Kennea2013}
Kennea, J. A., Burrows, D. N., Cummings, J., et al. 2013, The Astronomer¡¯s Telegram, 5124, 1

\bibitem{Li2014}
Li, L., Tong, H., Yan, W. M., et al. 2014, ApJ, 788, 16

\bibitem{Lynch2014}
Lynch, R. S., Archibald, R. F., Kaspi, V. M., Scholz, P. 2014, arXiv:1412.0610

\bibitem{Mori2013}
Mori, K., Gotthelf, E. V., Zhang, S., et al. 2013, ApJ, 770, L23

\bibitem{Rea2011}
Rea, N., \& Esposito, P. 2011, arXiv:1101.4472

\bibitem{Rea2013}
Rea, N., Esposito, P., Pons, J. A., et al. 2013, ApJ, 775, L34

\bibitem{Scholz2014}
Scholz P., Kaspi, V. M., \& Cumming, A. 2014, 786, 62

\bibitem{Shapiro (1983)}
Shapiro, S. L., \& Teukolsky S. A. 1983, Block holes, white dwarfs, and nuetron stars, John Wiley \& Sons, New York

\bibitem{TD (1996)}
Thompson, C., \& Duncan, R. C. 1996, ApJ, 473, 322

\bibitem{Thompson2000}
Thompson, C., Duncan, R. C., Woods, P. M., et al. 2000, ApJ, 543, 340

\bibitem{TLK (2002)}
Thompson, C., Lyutikov, M., \& Kulkarni, S. R. 2002, ApJ, 574, 332


\bibitem{Timokhin2008}
Timokhin, A. N., Eichler, D., \& Lyubarsky, Yu. 2008, ApJ, 680, 1398

\bibitem{Tong2010}
Tong, H., Xu, R. X., Peng, Q. H., Song, L. M. 2010, RAA, 10, 553

\bibitem{TongXu2013}
Tong, H., \& Xu, R. X. 2013, RAA, 13, 1207

\bibitem{Tong2013}
Tong, H., Xu, R. X., Song, L. M., \& Qiao, G. J. 2013, ApJ, 768, 144

\bibitem{Tong2014}
Tong, H. 2014, ApJ, 784, 86

\bibitem{TongXu2014}
Tong, H., \& Xu, R. X. 2014, AN, 335, 757

\bibitem{Vagino2013}
Vigano, D., Rea, N., Pons, J. A., et al. 2013, MNRAS, 434, 123

\bibitem{Watts2011}
Watts, A. L. 2011, arXiv:1111.0514

\bibitem{Woods2007}
Woods, P. M., Kouveliotou, C., Finger, M. H., et al. 2007, ApJ, 654, 470

\bibitem{Xu2001}
Xu, R. X., \& Qiao, G. J. 2001, ApJ, 561, L85

\bibitem{Younes2012}
Younes, G., Kouveliotou, C., Kargaltsev, O., et al. 2012, ApJ, 757, 39

\end{thebibliography}
\end{document}